\documentstyle[12pt]{article}

\begin{document}

\newcommand{\dfrac}[2]{\displaystyle{\frac{#1}{#2}}}

{\it University of Shizuoka}

\hspace*{9.5cm} {\bf US-96-09}\\[-.3in]


\hspace*{9.5cm} {\bf November 1996}\\[.3in]

\vspace*{.4in}

\begin{center}

{\large\bf  Flavor-Changing Neutral Currents Induced by }\\[.2in]

{\large\bf  the Democratic  Seesaw Mass Matrix  }\\[.3in]

{\bf Yoshio Koide}\footnote{
E-mail: koide@u-shizuoka-ken.ac.jp} \\

Department of Physics, University of Shizuoka \\ 
395 Yada, Shizuoka 422, Japan \\[.1in]

\vspace{.3in}

{\large\bf Abstract}\\[.1in]

\end{center}

\begin{quotation}
Flavor-changing neutral currents (FCNC)  are studied on the basis of a
``democratic seesaw" mass matrix model, which 
yields a singular enhancement of the top-quark mass $m_t$ 
and can give reasonable quark masses and CKM matrix elements. 
The most exciting aspect of the model is that the structure of the 
$6 \times 6$ right-handed fermion mixing matrix in the up-quark sector, 
$U^R_u$, shows an abnormal structure in contrast to that of $U^L_u$. 
This causes characteristic effects on the right-handed FCNC concerned 
with top quark. A single top-quark production at future 
$e^+ e^-$ colliders, $e^+ + e^- \rightarrow Z_R \rightarrow t + 
\overline{c}$ ($\overline{t}+c$), is discussed.
\end{quotation}

\newpage
\noindent
{\bf 1. Introduction}

\vglue.05in

Recently, in order to understand why the observed top-quark mass 
$m_t$ is so enhanced in contrast to the other quark masses, 
i.e., $m_t\gg m_b$, while $m_u \sim m_d$, 
Fusaoka and the author [1] have proposed a 
^^ ^^ democratic seesaw" mass matrix model for quarks and leptons 
$f_i \ (f=u, d, \nu, e; \ i=1, 2, 3)$. 
The $6\times 6$ mass matrix $M$ for the fermions ($f, F$) 
($F_i$ are hypothetical heavy fermions corresponding 
to the conventional quarks and leptons $f_i$) has the form [2]
$$
M=\left( \begin{array}{cc}
0 & m_L \\
m_R & M_F \\ 
\end{array} \right) 
= m_0\left( \begin{array}{cc}
0 & Z \\
\kappa Z & \lambda O_f \\ 
\end{array} \right) \ \ , 
\eqno(1.1)
$$
where the structure of the heavy fermion mass matrix $M_F$ has 
a form [3] 
[(unit matrix) $+$ (democratic matrix)] and is controlled by 
a $f$-dependent (complex) parameter $b_f e^{i\beta_f}$ as 
$$
O_f = {\bf 1} + 3 b_f e^{i\beta_f}X \ \ , 
\eqno(1.2)
$$
$$
{\bf 1} = \left(\begin{array}{ccc}
1 & 0 & 0 \\
0 & 1 & 0 \\
0 & 0 & 1 \\
\end{array} \right) \ \ , \ \ \ 
X = \frac{1}{3}\left(\begin{array}{ccc}
1 & 1 & 1 \\
1 & 1 & 1 \\
1 & 1 & 1 \\
\end{array} \right) \ \ , 
\eqno(1.3)
$$
while the matrix $m_0 Z = m_L = m_R/\kappa$ is universal 
for all fermion sectors ($f, F$), i.e., 
$$
m_0 Z = m_L = \frac{1}{\kappa}m_R = m_0 \left(\begin{array}{ccc}
z_1 & 0 & 0 \\
0 & z_2 & 0 \\
0 & 0 & z_3 \\ 
\end{array} \right) \ \ , 
\eqno(1.4)
$$
with $z_1^2 + z_2^2 + z_3^2=1$. 
For $\lambda^2 \gg \kappa^2 \gg 1$, the mass matrix (1.1) leads to 
the well-known ^^ ^^ seesaw" form [4] of the $3\times 3$ 
light-fermion mass matrix: 
$$
M_f \simeq m_LM_F^{-1} m_R = \frac{\kappa}{\lambda} m_0 Z O^{-1}_f Z \ \ .
\eqno(1.5)
$$
Note that the inverse matrix of $O_f$ defined in (1.2), $O_f^{-1}$, is 
given by 
$$
O_f^{-1} = {\bf 1} + 3a_f e^{i\alpha_f} X \ \ , 
\eqno(1.6)
$$
with
$$
a_f e^{i\alpha_f} = -\frac{b_f e^{i\beta_f}}{1 + 3b_f e^{i\beta_f}} \ \ , 
\eqno(1.7)
$$
so that the limit of $b_f e^{i\beta_f} \rightarrow -1/3$ leads to 
$a_f e^{i\alpha_f} \rightarrow \infty$. 
On the other hand, a democratic mass matrix [5] makes only one family heavy. 
Therefore, a choice $b_u = -1/3$ and $b_d \neq b_u$ (but $b_d\sim b_u$) 
can provide that $m_t \gg m_b$ with keeping $m_u \sim m_d$. 
By assuming that the parameter $b_f$ takes the value $b_e=0$
in the charged lepton sector, they have fixed the parameters $z_i$ as 
$$
\frac{z_1}{\sqrt{m_e}} = \frac{z_2}{\sqrt{m_\mu}} = \frac{z_3}{\sqrt{m_\tau}} 
= \frac{1}{\sqrt{m_e + m_\mu + m_\tau}} 
\eqno(1.8)
$$
from $M_e = m_0(\kappa/\lambda) Z\cdot{\bf 1}\cdot Z$.
Then, by choosing $b_u e^{i\beta_u} = -1/3$ and 
$b_d e^{i\beta_d}= -e^{- i\, 18^{\circ}}$ 
for the quark sectors together with $\kappa/\lambda = 0.02$, 
they have obtained reasonable quark mass ratios and 
Cabibbo-Kobayashi-Maskawa (CKM) [6] matrix parameters. 
Furthermore, recently, the author [7] has pointed out that 
the choice $b_f\simeq -1/2$ 
is favorable to understanding a large neutrino mixing which has been 
suggested from the atmospheric neutrino data [8].

Thus, the democratic seesaw mass matrix model can give favorable results 
in the phenomenology of the quark- and lepton-flavor physics, 
although the theoretical background 
of the model is still unclear: why the heavy fermion mass matrices 
$M_F$ take the form [(unit matrix)+(democratic matrix)]; 
what is the origin which yields $m_R=\kappa m_L$; 
why the nature chooses $b_u=-1/3$, $b_d\simeq -1$, $b_\nu\simeq -1/2$ 
and $b_e=0$; and so on.

The purpose of the present paper is not to answer these theoretical 
questions.  This democratic seesaw mass matrix model brings many 
interesting new aspects in the quark and lepton phenomenology, e.g., 
the predictions [1] of the considerably light mass of the fourth 
up-quark $t'$ ($\equiv u_4$) compared with the other heavy fermions, 
and so on. 
In order for the present model to be taken seriously, 
we needs more studies on the 
phenomenological characteristics of the model in contrast to the 
conventional mass matrix models.

As one of such characteristic features of the model, 
in the present paper, 
we will point out that the structure of the 
$6 \times 6$ right-handed fermion mixing matrix in the up-quark sector, 
$U^R_u$, takes a peculiar structure in contrast to that of 
$U^L_u$: 
For convenience, we denote the $6\times 6$ mixing matrix $U$ in 
terms of $3\times 3$ matrices $U_{ab}$ ($a,b=f,F$) 
$$
U=\left(
\begin{array}{cc}
U_{ff} & U_{fF} \\
U_{Ff} & U_{FF} \\
\end{array} \right) \ . 
\eqno(1.9)
$$
In a sector which satisfies the seesaw approximation (1.5), 
for example, for the down-quark sector, 
the mixing matrices $U_{fF}^L$ and $U_{Ff}^L$ 
($U_{fF}^R$ and $U_{Ff}^R$) are suppressed by a factor $1/\lambda$ 
($\kappa/\lambda$) compared with $U_{ff}^L=U_{ff}^{R*}$ and 
$U_{FF}^L=U_{FF}^{R*}$. 
However, in the up-quark sector, in which the seesaw approximation 
(1.5) is not valid any more, the mixing matrix elements 
$(U_{uU}^R)_{3i}$ and $(U_{Uu}^R)_{1i}$ $(i=1,2,3$) 
do not suffer such suppression, and,  instead, 
$(U_{uu}^R)_{3i}$ and $(U_{UU}^R)_{1i}$ are suppressed by a 
factor $\kappa/\lambda$ (see (3.8) later). 
This abnormal structure is due to the enhancement of the top-quark 
mass $m_t\equiv m_3^u$ (and the suppression of the fourth up-quark 
mass $m_4^u$) as stated in Sec.3.
This will cause characteristic effects on the right-handed 
flavor-changing neutral currents (FCNC) concerned with top quark. 

In Sec.2, we give an outline of the model. 
In Sec.3, we give the $6\times 6$ mixing matrices $U^L_f$ and 
$U^R_f$, and in Sec.4, we give the induced right-handed FCNC structure.
In Sec.5, as an example of the observable effects of the FCNC, 
a single top-quark production in future $e^+ e^-$ colliders is 
discussed.

\vglue.2in
\noindent
{\bf 2. Outline of the model}

\vglue.05in

In the present model, quarks and leptons $f_i$ belong to 
$f_L = (2,1)$ and 
$f_R = (1, 2)$ of SU(2)$_L \times $SU(2)$_R$ 
and heavy fermions $F_i$ are vector-like, i.e., 
$F_L = (1, 1)$ and $F_R = (1, 1)$. 
The vector-like fermions $F_i$ acquire masses $M_F$ at 
an energy scale of the order $\lambda m_0$. Note that in 
our model, there is no Higgs boson with (2,2) of 
SU(2)$_L \times$SU(2)$_R$ 
differently from the standard SU(2)$_L \times $SU(2)$_R$ model [9].
The SU(2)$_L$ and SU(2)$_R$ symmetries are broken by 
Higgs bosons $\phi_L=(\phi_L^+, \phi_L^0)$ and 
$\phi_R=(\phi_R^+, \phi_R^0)$, which belong to (2,1) and (1,2) of 
SU(2)$_L \times $SU(2)$_R$, respectively.
We assume that these Higgs bosons 
couple to the fermions universally:
$$
H_{Yukawa}=\sum_{i=1}^3  y_{Li}\left[
(\overline{u} \ \overline{d})_{L i} \left(
\begin{array}{c}
\phi_L^+ \\
\phi_L^0 
\end{array} \right) D_{Ri} +   
 (\overline{u} \ \overline{d})_{L i} \left(
\begin{array}{c} \overline{\phi}_L^0 \\
-\phi_L^- 
\end{array}
\right) U_{Ri} \right]
$$
$$
+ h.c. +(L\leftrightarrow R) + [(u,d,U,D)\rightarrow (\nu,e,N,E)]
\ , \eqno(2.1)   
$$
where $y_{Li}$ and $y_{Ri}$ are real parameters, and they are 
universal for all the fermion sectors.
Therefore, the mass matrix  which is sandwiched by 
$(\overline{f}_L, \overline{F}_L)$  and $(f_R, F_R)^T$ is given by
the $6\times 6$ matrix  (1.1), 
where $m_L=y_{L_i}\langle \phi_L^0\rangle $.

As seen in Ref.[1], 
phenomenologically, up- and down-quark masses are well described by 
choosing $b_u=-1/3$ and $b_d \simeq -1$:
$$
m_u \simeq \frac{3 m_e}{4 m_\tau} \frac{\kappa}{\lambda}m_0 \ \ , \ \ 
m_c \simeq 2 \frac{m_\mu}{m_\tau} \frac{\kappa}{\lambda}m_0 \ \ , \ \ 
m_t \simeq \frac{1}{\sqrt{3}} m_0 \ \ , 
$$
$$
m_4^u \simeq \frac{1}{\sqrt{3}}\kappa m_0 \ \ , \ \ 
m_5^u \simeq m_6^u \simeq \lambda m_0 \ \ , 
\eqno(2.2)
$$
$$
m_d \simeq \frac{1}{2|\sin(\beta_d/2)|} \frac{m_e}{m_\tau} 
\frac{\kappa}{\lambda}m_0 \ \ , \ \ 
m_s \simeq 2\left|\sin\frac{\beta_d}{2}\right|\frac{m_\mu}{m_\tau} 
\frac{\kappa}{\lambda}m_0 \ \ , \ \ 
m_b \simeq \frac{1}{2} \frac{\kappa}{\lambda}m_0 \ \ , 
$$
$$
m_4^d \simeq m_5^d \simeq \lambda m_0 \ \ , \ \ 
m_6^d \simeq 2\sqrt{1 + 3\sin^2(\beta_d/2)}\lambda m_0 \ \ . 
\eqno(2.3)
$$
The observed quark mass ratios $m_c/m_t$ and $m_d/m_s$ require 
$\kappa/\lambda \simeq 0.02$ and $|\beta_d|\simeq 18^{\circ}$, 
respectively.
These input values can give reasonable values of the CKM mixings 
$|V_{ij}|$.

Note that only the fourth up-quark mass $m_4^u$ is remarkably light 
compared other heavy fermions. 
The enhancement of the top-quark mass $m_t \ (\equiv m_3^u)$ is caused at 
the cost of the lightening of $u_4 \ (\equiv U_1)$. 
We speculate 
${m_4^u}/{m_3^u} \simeq \kappa \sim {m_{W_R}}/{m_{W_L}}$, i.e., 
$m_4^u \sim 10^3$ GeV. 
We can expect the observation of the fourth up-quark $u_4$ at 
an energy scale at which we can observe the right-handed weak bosons 
$W_R$.

\vglue.2in
\noindent
{\bf 3. Peculiar structure of the mixing matrix $U^R_u$}

\vglue.05in

In the Ref.[1], of the $6\times 6$ mixing matrices, only the 
left-handed light-fermion-mixing part $U_{ff}^L$ have been studied. 
In the present paper, we investigate the $6\times 6$ mixing matrix 
(1.9) and will find that right-handed mixing matrix 
$U^R_u$ for the up-quark sector $(u,U)$ has a peculiar structure 
in contrast to $U^L_u$, although the mixing matrix $U^R_d$ 
for the down-quark sector $(d,D)$ has a similar structure to 
$U^L_d$.

For the case where the seesaw expression (1.1) is a good approximation, 
the $6 \times 6$ mixing matrices $U^L$ and $U^R$ for the fermions 
$(f_L, F_L)$ and $(f_R, F_R)$ are given by 
$$
U^L \simeq \left(\begin{array}{cc}
U^L_f & 0 \\ 
0 & U^L_F \\ 
\end{array} \right) 
\left(\begin{array}{cc}
{\bf 1} & -m_L M_F^{-1} \\ 
M_F^{\dagger -1} m_L & {\bf 1} \\ 
\end{array} \right) 
= \left( \begin{array}{cc}
U^L_f & -U^L_f m_L M_F^{-1} \\ 
U^L_F M_F^{\dagger -1} m_L & U^L_F \\ 
\end{array} \right) \ \ , 
\eqno(3.1)
$$
$$
U^R \simeq \left(\begin{array}{cc}
U^R_f & 0 \\ 
0 & U^R_F \\ 
\end{array} \right) 
\left(\begin{array}{cc}
{\bf 1} & -m_R M_F^{\dagger -1} \\ 
M_F^{-1} m_R & {\bf 1} \\ 
\end{array} \right) 
= \left(\begin{array}{cc}
U^R_f & -U^R_f m_R M_F^{\dagger -1} \\ 
U^R_F M_F^{-1} m_R & U^R_F \\ 
\end{array} \right) \ \ , 
\eqno(3.2)
$$
where $U_f^L$, $U_f^R$, $U_F^L$ and $U_F^R$ are defined by
$$
-U^L_f m_L M_F^{-1} m_R U_f^{R\dagger} = D_f \ \ , 
\eqno(3.3)
$$
$$
U^L_F M_F U_F^{R\dagger} = D_F \ \ , 
\eqno(3.4)
$$
$D_f = {\rm diag}(m_1^f, m_2^f, m_3^f)$ and 
$D_F = {\rm diag}(m_4^f, m_5^f, m_6^f) \equiv 
{\rm diag}(m_1^F, m_2^F, m_3^F)$. 
The mixing matrix $U^R$ is related to $U^L$ as follows:
$$
\begin{array}{ll}
(U^R_{ff})_{ij} = (U^L_{ff})_{ij}^*  \ , \ \ & 
(U^R_{FF})_{ij} = (U^L_{FF})_{ij}^*  \ , \\
(U^R_{fF})_{ij} = \kappa (U^L_{fF})_{ij}^*  \ , \ \ &
(U^R_{Ff})_{ij} = \kappa (U^L_{Ff})^*_{ij} \ ,  
\end{array}
\eqno(3.5)
$$
where $i= 1, 2, 3$. 
For example, the explicit numerical result of $U_d^L$ 
without the seesaw approximation is given by
$$
|U^L_d| = \left(
\begin{array}{lll|lll} 
0.9772 &  0.2061  &  0.0506 &  0.0490\frac{1}{\lambda} 
& 0.0007\frac{1}{\lambda} & 4\times 10^{-5}\frac{1}{\lambda} \\
 0.2117 & 0.9540  & 0.2124 & 0.2063\frac{1}{\lambda} 
&  0.0646\frac{1}{\lambda} &  0.0035\frac{1}{\lambda} \\
0.0137 & 0.2179  & 0.9759 & 0.4335\frac{1}{\lambda} 
& 0.4809\frac{1}{\lambda} &  0.5251\frac{1}{\lambda} \\ 
\hline
0.0118\frac{1}{\lambda} & 0.1649\frac{1}{\lambda} & 0.0209\frac{1}{\lambda} 
& 0.7176 & 0.6961 & 0.0215 \\
0.0064\frac{1}{\lambda} & 0.1011\frac{1}{\lambda} & 0.7927\frac{1}{\lambda} 
& 0.3895 & 0.4268 & 0.8162 \\
0.0046\frac{1}{\lambda} & 0.0660\frac{1}{\lambda} & 0.2706\frac{1}{\lambda} 
& 0.5773 & 0.5773 & 0.5774 
\end{array} \right) \ ,
\eqno(3.6)
$$
where we have used the input values $b_d=-1$, $\beta_d=-18^\circ$,
$\kappa/\lambda=0.02$ and $\kappa=10$ according to Ref.~[1].

However, for the up-quark sector with $b_f e^{i\beta_f}=-1/3$, 
the relations (3.5) are not valid any longer. 
For up-quark sector, the $6 \times 6$ left-handed mixing matrix $U^L$ 
is given by 
$$
U^L_u = \left(
\begin{array}{lll|lll} 
+0.9994 & -0.0349 & -0.0084 & -0.0247 \frac{1}{\lambda} & 
+6\times 10^{-5}\frac{1}{\lambda} & +4\times 10^{-6}\frac{1}{\lambda} \\
+0.0319 & +0.9709 & -0.2373 & -0.2051\frac{1}{\lambda} 
& -0.4346\frac{1}{\lambda} & +0.0259\frac{1}{\lambda} \\
+0.0165 & +0.2369 & +0.9714 & +0.8990\frac{1}{\lambda} & 
+0.8431\frac{1}{\lambda} & -0.0444\frac{1}{\lambda} \\ 
\hline
+0.0934\frac{1}{\lambda} & +0.1114\frac{1}{\lambda} 
& -1.0365\frac{1}{\lambda} & +0.5774 & +0.5774 & +0.5772 \\
-0.0118\frac{1}{\lambda} & +0.1649\frac{1}{\lambda} 
& +0.0209\frac{1}{\lambda} & -0.7176 & +0.6961 & +0.0215 \\
-0.0064\frac{1}{\lambda} & -0.1011\frac{1}{\lambda} 
& +0.7927\frac{1}{\lambda} & -0.3894 & -0.4267 & +0.8163 
\end{array} \right) \ ,
\eqno(3.7)
$$
while the right-handed mixing matrix $U^R_u$ is given by 
$$
U^R_u = \left(
\begin{array}{lll|lll} 
+0.9994 & -0.0349 & -0.0084 & -0.0247\frac{\kappa}{\lambda} & 
+6\times 10^{-5}\frac{\kappa}{\lambda} 
& +4\times 10^{-6}\frac{\kappa}{\lambda} \\
+0.0319 & +0.9709 & -0.2373 & -0.2051\frac{\kappa}{\lambda} 
& -0.4346\frac{\kappa}{\lambda} & +0.0259\frac{\kappa}{\lambda} \\
+0.0256\frac{\kappa}{\lambda} & +0.3459\frac{\kappa}{\lambda} 
& -0.0747\frac{\kappa}{\lambda} & +0.5773 & +0.5773 & +0.5774 \\
\hline
+0.0165 & +0.2369 & +0.9713 & +0.3274\frac{\kappa}{\lambda} 
& +0.2716\frac{\kappa}{\lambda} & -0.6160\frac{\kappa}{\lambda} \\ 
-0.0118\frac{\kappa}{\lambda} & +0.1649\frac{\kappa}{\lambda} 
& +0.0209\frac{\kappa}{\lambda} & -0.7176 & +0.6961 & +0.0215 \\
-0.0064\frac{\kappa}{\lambda} & -0.1011\frac{\kappa}{\lambda} 
& +0.7929\frac{\kappa}{\lambda} & -0.3894 & -0.4267 & +0.8161 
\end{array} \right) \ .
\eqno(3.8)
$$

Note that $U^R_u$, (3.8), shows a peculiar structure as if the third and 
fourth rows of the matrix $U^R_u$ are exchanged each other in 
contrast to the mixing matrices (3.6) and (3.7). 
As seen in Sec.~4, for the FCNC phenomenology, the structure of 
$U^R_{uU}$ plays an essential role. 
The $3\times 3$ matrix $U^R_{uU}$ is analytically given  by 
\renewcommand{\arraystretch}{2}
$$
U^R_{uU}\simeq \left(
\begin{array}{ccc}
 -\dfrac{3}{2}z_1 \dfrac{\kappa}{\lambda} & 
\dfrac{3}{4}\dfrac{z_1^3}{z_2^2} \dfrac{\kappa}{\lambda} & 
\dfrac{3}{4}\dfrac{z_1^3}{z_3^2} \dfrac{\kappa}{\lambda} \\
- z_1 \dfrac{\kappa}{\lambda} &
-2 z_2 \dfrac{\kappa}{\lambda} &
2\dfrac{z_2^3}{z_3^2} \dfrac{\kappa}{\lambda} \\
\dfrac{1}{\sqrt{3}} & \dfrac{1}{\sqrt{3}} & \dfrac{1}{\sqrt{3}} \\
\end{array} \right) \ .
\eqno(3.9)
$$

\renewcommand{\arraystretch}{1}
This peculiar structure can be understood from the following situation. 
For convenience, we transform the heavy fermion basis in which 
$M_F=m_0 \lambda O_f$ is given by (1.2) 
into a basis in which the heavy fermion mass matrix $M_F$ is 
diagonalized as 
$$
A M_F A^{-1} = \widetilde{M}_F= m_0 \lambda \, 
{\rm diag}(1+3b_fe^{i\beta_f},1,1) \ . 
\eqno(3.10)
$$ 
Then, the $6 \times 6$ mass matrix for up-quark sector is transformed 
as 
$$
\widetilde{M} = \left(\begin{array}{cc}
{\bf 1} & 0 \\ 
0 & A \\ 
\end{array} \right) M 
\left(\begin{array}{cc}
{\bf 1} & 0 \\ 
0 & A^T \\ 
\end{array} \right) = 
\left(\begin{array}{cc}
0 & \widetilde{m}_L \\ 
\widetilde{m}_R & \widetilde{M}_U \\ 
\end{array} \right) = 
m_0 \left(\begin{array}{cc}
0 & \widetilde{Z}^T \\ 
\kappa\widetilde{Z} & \lambda\widetilde{O}_u \\ 
\end{array} \right) \ \ , 
\eqno(3.11)
$$
where 
$$
\renewcommand{\arraystretch}{2}
\widetilde{Z} = AZ = \left(
\begin{array}{ccc}
\dfrac{1}{\sqrt{3}} z_1 & \dfrac{1}{\sqrt{3}} z_2 
& \dfrac{1}{\sqrt{3}} z_3 \\
-\dfrac{1}{\sqrt{2}} z_1 & \dfrac{1}{\sqrt{2}} z_2 & 0 \\
-\dfrac{1}{\sqrt{6}} z_1 & -\dfrac{1}{\sqrt{6}} z_2 & 
\dfrac{2}{\sqrt{6}} z_3 
\end{array} \right) \ , 
\eqno(3.12)
\renewcommand{\arraystretch}{1}
$$ 
$$
\widetilde{O}_u = A O_u A^T = {\rm diag}(0, 1, 1) \ \ . 
\eqno(3.13)
$$
The mixing matrices $U^L$ and $U^R$ are given as matrices which 
diagonalize Hermitian matrices 
$H_L \equiv \widetilde{M}\widetilde{M}^\dagger$ 
and $H_R \equiv \widetilde{M}^\dagger\widetilde{M}$, 
respectively:
$$
H_L = \widetilde{M}\widetilde{M}^\dagger = 
m_0^2 \left(\begin{array}{cc}
\widetilde{Z}^T \widetilde{Z} & \lambda\widetilde{Z}^T\widetilde{O}_u \\ 
\lambda\widetilde{O}_u\widetilde{Z} & 
\lambda^2\widetilde{O}_u^2 + \kappa^2 \widetilde{Z}\widetilde{Z}^T \\ 
\end{array} \right) \ \ , 
\eqno(3.14)
$$
$$
H_R = \widetilde{M}^\dagger\widetilde{M} = m_0^2 
\left(\begin{array}{cc}
\kappa^2\widetilde{Z}^T\widetilde{Z} 
& \kappa \lambda \widetilde{Z}^T \widetilde{O}_u \\ 
\kappa\lambda \widetilde{O}_u\widetilde{Z} & 
\lambda^2 \widetilde{O}_u^2 + \widetilde{Z}\widetilde{Z}^T \\ 
\end{array} \right) \ \ . 
\eqno(3.15)
$$
In our scheme, the numbering of the fermions $f_i$  
($i= 1, 2, \cdots, 6$) 
is defined as $m_1^f < m_2^f < m_3^f < m_4^f < 
m_5^f < m_6^f$. 
Since $(\widetilde{O}_u^2)_{11}=0$, we see that 
$(H_L)_{33}=m_0^2(\widetilde{Z}^T\widetilde{Z})_{33}$ and 
$(H_L)_{44}=m_0^2\kappa^2 (\widetilde{Z}\widetilde{Z}^T)_{11}$, i.e.,
$(H_L)_{33}\ll (H_L)_{44}$, 
while $(H_R)_{33} = m_0^2\kappa^2(\widetilde{Z}^T\widetilde{Z})_{33}$ and 
$(H_R)_{44} = m_0^2 (\widetilde{Z}\widetilde{Z}^T)_{11}$, i.e., 
$(H_R)_{33}\gg (H_R)_{44}$,. 
This causes the exchange $U_{3i} \leftrightarrow U_{4i}$. 
On the other hand, for the ordinary case $b_f e^{i\beta_f} \neq -1/3$, 
both $(H_L)_{44}$ and $(H_R)_{44}$ are of the order of $m_0^2\lambda^2$, 
i.e., $(H_L)_{33}\ll (H_L)_{44}$ and $(H_R)_{33}\ll (H_R)_{44}$,
so that such an exchange $U_{3i} \leftrightarrow U_{4i}$ is not caused.

As a result, the $6\times 6$ mixing matrix $V^R$ 
for the right-handed charged 
currents shows an abnormal structure in contrast to the CKM mixing 
matrix $V^L$: the magnitudes of 
$V^R_{tq} \ (q=d, s, b)$ are suppressed by the order of $\kappa/\lambda$, 
while the magnitudes of $V^R_{t'q} \ (t' \equiv u_4)$ are given by 
$$
|V^R_{t'q}| \simeq |V^L_{tq}| \ \ , \ (q = d, s, b)  \ , 
\eqno(3.16)
$$
If we suppose $\kappa \sim 10$, the mass of the fourth up-quark $t'$ 
is of the order of $10^3$ GeV, so that 
we can expect observation of 
a single $t'$-production via $W_R$-exchange, $u+d \rightarrow d+t'$, 
at LHC, because of $|V_{ud}^R| \simeq 1$ and $|V_{t'd}^R| \sim 10^{-2}$.

\vglue.2in
\noindent
{\bf 4. Structure of FCNC}

\vglue.05in

When the mass matrix $M$ given in (1.1) is transformed as 
$$
\overline{\psi_L} M \psi_R + h.c. = \overline{\psi'_L}D\overline{\psi'_R} + 
h.c. \ \ , 
\eqno(4.1)
$$
where $\psi = (f, F)^T$, and $\psi' = U\psi$ is the mass-eigenstates, 
the vertex $\overline{\psi}_A\Gamma^{AB}\psi_B$  ($A, B= L, R$) is also 
transformed into $\overline{\psi}'_A\Gamma'^{AB}\psi'_B$, 
where 
$$
\Gamma'^{AB} = U_A \Gamma^{AB} U_B^\dagger \ \ . 
\eqno(4.2)
$$
For simplicity, hereafter, we drop the indices $A, \ B$.
Correspondingly to (1.9), we denote the $6 \times 6$ matrix $\Gamma$ 
in terms of $3 \times 3$ matrices  $\Gamma_{ab}$  ($a, b=f,F)$ as 
$$
\Gamma = \left(\begin{array}{cc}
\Gamma_{ff} & \Gamma_{fF} \\ 
\Gamma_{Ff} & \Gamma_{FF} \\ 
\end{array} \right)  \ . 
\eqno(4.3)
$$
Our interest is in the physical vertex $\Gamma'_{ff}$ which is 
given by 
$$
\Gamma'_{ff} = \sum_{a} \sum_{b} 
U_{fa}\Gamma_{ab} U_{fb}^\dagger \ \ , 
\eqno(4.4)
$$
where $U_{ab}^\dagger \equiv (U_{ab})^\dagger = (U^\dagger)_{ba}$, 
because $(\Gamma'_{ff})_{ij}$ with $i \neq j$ mean transitions 
between $f_i$ and $f_j$, i.e.,  appearance of the FCNC. 

In our SU(2)$_L \times $SU(2)$_R \times $U(1)$_Y$ gauge model, 
the neutral currents  $J_L^\mu = g_L^Z \overline{\psi}\Gamma_L^\mu \psi$, 
which couple with the left-handed weak boson $Z_L^\mu$, 
are given by 
$$
\Gamma_L^\mu = \left(\begin{array}{cc}
c_L^f{\bf 1} & 0 \\ 
0 & c_L^F{\bf 1} \\ 
\end{array} \right) \cdot \frac{1}{2} \gamma^\mu (1-\gamma_5) + 
\left(\begin{array}{cc}
d_L^f{\bf 1} & 0 \\ 
0 & d_L^F{\bf 1} \\ 
\end{array} \right) \cdot \frac{1}{2} \gamma^\mu (1+ \gamma_5) \ \ , 
\eqno(4.5)
$$
where 
$$
\begin{array}{rll}
c_L^f & = \pm \frac{1}{2} & - \sin^2\theta_L Q_f \ \ , \\ 
c_L^F & =  & -\sin^2\theta_L Q_F \ \ , \\ 
\end{array} 
\eqno(4.6)
$$
$$
\begin{array}{rll}
d_L^f & = \pm\frac{1}{2}h_{L} & -\sin^2\theta_L Q_f \ \ , \\ 
d_L^F & =  & - \sin^2\theta_L Q_F \ \ , \\ 
\end{array} 
\eqno(4.7)
$$
$$
\sin^2\theta_L = 1 - m_{W_L}^2/m_{Z_L}^2 \ \ , 
\eqno(4.8)
$$
$$
h_{L} = -\frac{\sin^2\theta_L}{1 - \varepsilon/\cos^2\theta_L} 
\frac{\varepsilon}{\cos^2\theta_L} \ \ , 
\eqno(4.9)
$$
$$
\varepsilon = m_{W_L}^2/m_{W_R}^2 \ \ , 
\eqno(4.10)
$$
the factor $\pm\frac{1}{2}$ takes $+\frac{1}{2}$ and $-\frac{1}{2}$ for 
up- and down-fermions, respectively, and $Q_f \ (Q_F)$ is 
charge of the fermion $f$  $(F)$. 
Using the unitary condition for $U_{ab}$, $U_{ff}U_{ff}^\dagger + 
U_{fF}U_{fF}^\dagger = {\bf 1}$, we can express the physical vertex 
$\Gamma'_{Lff}$ as
$$
\begin{array}{c}
\Gamma^{'\mu}_{Lff} = \left(c_L^f U_{ff}^LU_{ff}^{L\dagger} + 
c_L^F U_{fF}^LU_{fF}^{L\dagger}\right) 
\cdot \frac{1}{2}\gamma^\mu(1 - \gamma_5) \\ 
+ \left(d_L^f U_{ff}^RU_{ff}^{R\dagger} + 
d_L^F U_{fF}^RU_{fF}^{R\dagger}\right) \cdot 
\frac{1}{2}\gamma^\mu(1 + \gamma_5) \\ 
= \left[c_L^f {\bf 1} - (c_L^f - c_L^F)U_{fF}^LU_{fF}^{L\dagger}
\right] \cdot 
\frac{1}{2}\gamma^\mu(1-\gamma_5) \\ 
+ \left[d_L^f {\bf 1} - (d_L^f - d_L^F)U_{fF}^RU_{fF}^{R\dagger}
\right] \cdot 
\frac{1}{2}\gamma^\mu (1 + \gamma_5) \ \ . \\ 
\end{array} \eqno(4.11)
$$
Similarly, for the neutral current $J_R^\mu = g_R^Z \overline{\psi'} 
\Gamma_R^{'\mu} \psi'$, which couples with the right-handed weak boson 
$Z_L$, we obtain 
$$
\begin{array}{c}
\Gamma^{'\mu}_{Rff} = \left[c_R^f{\bf 1} - 
(c_R^f - c_R^F)U_{fF}^RU_{fF}^{R\dagger}\right] 
\cdot \frac{1}{2} \gamma^\mu(1 + \gamma_5) \\ 
+ \left[d_R^f {\bf 1} - (d_R^f - d_R^F) U_{fF}^LU_{fF}^{L\dagger}
\right] \cdot 
\frac{1}{2}\gamma^\mu(1-\gamma_5) \ \ , \\ 
\end{array} \eqno(4.12)
$$
where 
$$
\begin{array}{rll}
c_R^f & = \pm \frac{1}{2} & - \sin^2\theta_R Q_f \ \ , \\ 
c_R^F & =  & - \sin^2\theta_RQ_F \ \ , \\ 
\end{array} \eqno(4.13)
$$
$$
\begin{array}{rll}
d_R^f & = \pm \frac{1}{2}h_{R} & -\sin^2\theta_R Q_f \ \ , \\
d_R^F & =  & -\sin^2\theta_RQ_F \ \ , \\
\end{array} \eqno(4.14)
$$
$$
\sin^2\theta_R = 1 - m_{W_R}^2/m_{Z_R}^2 \ \ , 
\eqno(4.15)
$$
$$
h_{R} = -\frac{\sin^2\theta_R}{1 - \varepsilon\cos^2\theta_R} \ \ , 
\eqno(4.16)
$$
$$
g_R^Z = -g_L^Z\frac{\sin\theta_L}{\sin\theta_R\cos\theta_R} 
\sqrt{\frac{1 - \varepsilon\cos^2\theta_R}
{1 - \varepsilon/\cos^2\theta_L}}
$$
$$
=\frac{e}{\cos\theta_L\sin\theta_R\cos\theta_R} 
\sqrt{\frac{1 - \varepsilon\cos^2\theta_R}
{1 - \varepsilon\cos^2\theta_R/\cos^2\theta_L}} \ \ . 
\eqno(4.17)
$$

Note that the FCNC are induced by the second terms 
$U_{fF}U_{fF}^\dagger$ 
with magnitude $(c^f - c^F)$ $[(d^f - d^F)]$. 
The numerical results of $\xi^L\equiv U_{fF}^LU_{fF}^{L\dagger}$ 
and $\xi^R\equiv U_{fF}^RU_{fF}^{R\dagger}$ are as follows:
$$
\xi_u^L = \left(\begin{array}{ccc}
2.43 \times 10^{-9} & 2.01 \times 10^{-8} & -8.85 \times 10^{-8} \\
2.01 \times 10^{-8} & 9.26 \times 10^{-7} & -2.21 \times 10^{-6} \\
-8.85 \times 10^{-8} & -2.21 \times 10^{-6} & 
6.08 \times 10^{-6} 
\end{array} \right) \ \ , 
\eqno(4.18)
$$
$$
\xi_u^R = \left( \begin{array}{ccc}
2.43 \times 10^{-7} & 2.01 \times 10^{-6} & -2.84 \times 10^{-4} \\
2.01 \times 10^{-6} & 9.26 \times 10^{-5} & -7.09 \times 10^{-3} \\
-2.84 \times 10^{-4} & -7.09 \times 10^{-3} & 
1.000  
\end{array} \right) \ \ , \eqno(4.19)
$$
$$
|\xi_d^L| =| \xi_d^R| 
= \left(\begin{array}{ccc}
9.61 \times 10^{-9} & 4.03 \times 10^{-8} & 8.52 \times 10^{-8} \\
4.03 \times 10^{-8} & 1.87 \times 10^{-7} & 3.51 \times 10^{-7} \\
8.52 \times 10^{-8} & 3.51 \times 10^{-7} & 2.78 \times 10^{-6} 
\end{array} \right) \ \ , \eqno(4.20)
$$
where, for simplicity, for $\xi_d$, we have denoted only the magnitudes. 

As seen from (4.18)--(4.20), the magnitudes of 
$(\xi^L_u)_{12}$ and $(\xi_d^L)_{12}$ 
are sufficiently small, so that the contributions to 
$D^0\overline{D}^0$ and 
$K^0\overline{K}^0$ mixings and the rare decays of 
$D^0$ and $K^0$ are safely negligible. 
Although the value $(\xi^R_u)_{33} \simeq 1$ is noticeable, 
it is hard to observe the effects. 

We give attention to the magnitudes 
$\xi^R_{ut}\equiv|(\xi^R_u)_{13}|= 2.8 \times 10^{-4}$ and 
$\xi^R_{ct}\equiv|(\xi^R_u)_{23}| = 0.0071$, 
which are considerably large compared with the other 
off-diagonal elements. 
The mixing $u\leftrightarrow t$ can contribute to
a single top-quark production, 
$e^- + u \rightarrow e^- + t$, at HERA. 
However, because of the smallness of the value 
$\xi^R_{ut}$, the cross section 
$\sigma(e^-+p\rightarrow e^-+t+X)$ is of the order of 
$10^{-8}$ pb for $m_{Z_R}\simeq 0.9$ TeV, so that 
it is hard to observe the single top-quark production at HERA.
On the other hand, the mixing $c\leftrightarrow t$ can contribute to 
a single top-quark production, 
$e^+ + e^- \rightarrow Z_R \rightarrow t + \overline{c}$
 $(\overline{t} + c)$, 
at super $e^+e^-$ colliders. 
In the next section, we will discuss a possibility of the 
observation of $e^+ + e^- \rightarrow t + \overline{c}$ 
($c+\overline{t}$).

\vglue.2in
\noindent
{\bf 5. Single top-quark production at future $e^+ e^-$ colliders}

\vglue.05in

The matrix element of the reaction 
$e^+ + e^- \rightarrow t+ \overline{c}$ is given by 
$$
{\cal M} = G_L (\overline{u}_e\gamma_\mu (v_L -a_L \gamma_5)v_e)
\frac{g^{\mu\nu}-q^\mu q^\nu/m_{Z_L}^2}{
q^2-m_{Z_L}^2 +im_{Z_L}\Gamma_{Z_L} } 
(\overline{u}_t\gamma_\nu (1+\gamma_5)v_c) 
$$
$$
+ (L\rightarrow R) \ , \eqno(5.1)
$$
where we have neglected the $\xi_{ct}^L$-term and 
$$
G_L=-\frac{1}{2}\varepsilon \tan^2\theta_L (g_L^Z)^2 \xi^R_{ct} \ , \ \ 
\ G_R=\frac{1}{2} (g_R^Z)^2 \xi^R_{ct} \ , 
\eqno(5.2)
$$
$$
\begin{array}{l}
v_L=\frac{1}{4}(1-4\sin^2\theta_L) \ , \ \ \ 
a_L= \frac{1}{4} \\
v_R=\frac{1}{4}(1-5\sin^2\theta_R) \ , \ \ \ 
a_R= \frac{1}{4}(1+\sin^2\theta_R) \ , \\
\end{array}
\eqno(5.3)
$$
for $|\varepsilon|\ll 1$.
For $\sin^2\theta_R \sim \sin^2\theta_L$, the second term in (5.1), 
the $Z_R$-exchange term, is dominated.
When we neglect the contribution of the $Z_L$-exchange term, we obtain
$$
\sigma(e^+ e^-\rightarrow t\overline{c}) \simeq G_R^2 
\frac{v_R^2+a_R^2}{2\pi}\frac{(s-m_t^2)^2(2s+m_t^2)}{
s^2[(s-m_{Z_R}^2)^2+(m_{Z_R}\Gamma_{Z_R})^2]} \ .
\eqno(5.4)
$$
Here, the decay width $\Gamma_{Z_R}$ is given by
$$
\Gamma_{Z_R}=\frac{(g_R^Z)^2}{12\pi} \sum_f (v_R^2+a_R^2)m_{Z_R} \ , 
\eqno(5.5)
$$
where the sum is taken over all quarks and leptons, so that 
$\sum (v_R^2 +a_R^2)=3-3\sin^2\theta_R +2\sin^4\theta_R$. 
In order to give rough estimates, we take the inputs 
$\sin^2\theta_R=\sin^2\theta_L=0.23$, 
$\alpha^{-1}=120$, $m_{Z_R}=10 m_{Z_L}=0.9$ TeV, and 
$m_t=0.18$ TeV, and we obtain $\Gamma_{Z_R}=0.049 m_{Z_R}$ and 
$$
\begin{array}{ll}
\sigma =6.0\times 10^{-7}\ {\rm pb} & 
{\rm at}\ \sqrt{s}=0.2 \ {\rm TeV} \ , \\
\sigma =3.1\times 10^{-5}\ {\rm pb} & 
{\rm at}\ \sqrt{s}=2m_t=0.36 \ {\rm TeV} \ , \\
\sigma =1.1\times 10^{-4}\ {\rm pb} & 
{\rm at}\ \sqrt{s}=0.5 \ {\rm TeV} \ , \\
\sigma =7.5\times 10^{-4}\ {\rm pb} & 
{\rm at}\ \sqrt{s}=0.7 \ {\rm TeV} \ , \\
\end{array}
\eqno(5.6)
$$
where $\sigma=\sigma(t\overline{c})+\sigma(c\overline{t})$. 
The value of $\sigma$ is highly dependent on the choice of the value $m_{Z_R}$ 
(the value of $\sigma$ is roughly proportional to $m_{Z_R}^{-4}$). 
For example, if we take $m_{Z_R}=0.5$ TeV, 
the values of $\sigma$ given in (5.6) 
become large by a factor 11 times. 
Therefore, the values given in (5.6) should not be taken rigidly.
However, even taking such  ambiguity into consideration, 
the value of $\sigma$ at $\sqrt{s}=0.2$ TeV is 
too small to observe the single top-quark production at LEP. 

In order to contrast the $t+\overline{c}$ production to 
the ordinary $t+\overline{t}$ 
production, it is favorable that the observation of 
$t+\overline{c}$ is carry out at an $e^+e^-$ 
energy $\sqrt{s}$ which is  slightly smaller than $\sqrt{s}=2m_t$.
However, even if we have an $e^+ e^-$ collider with $L=10^{34}$ 
cm$^{-2}$s$^{-1}$, the value of $\sigma L$ is, at most, 
$\sigma L=0.026$ day$^{-1}$ at $\sqrt{s}=2m_t$, 
so that it is not so easy to detect the single top-quark production.
For example, a future $e^+ e^-$ collider JLC is designed as 
$L=8\times 10^{33}$ cm$^{-2}$s$^{-1}$ 
at $\sqrt{s}=0.5$ TeV in JLC [10]. 
This collider parameter gives $\sigma L=0.078$ day$^{-1}$ 
(i.e., one event/two weeks). 
Therefore, the single top-quark production will be barely detectable 
at such a future collider. 

If we can observe the direct production of $Z_R$ at future $e^+ e^-$ 
colliders, then we will reach the observation of the single top 
quark production at $\sqrt{s}=m_{Z_R}$: for example, 
for $m_{Z_R}=0.9$ TeV, we obtain $\sigma=0.085$ pb, 
which gives $\sigma L=3.9$ hour$^{-1}$ for $L=1.26\times 10^{34}$ 
cm$^{-2}$s$^{-1}$ at JLC [10].

\vglue.2in
\noindent
{\bf 6. Summary}

\vglue.05in

In conclusion, we have pointed out that 
the right-handed flavor-mixing matrix in the democratic 
seesaw mass matrix model takes a peculiar structure for the up-quark 
sector.
The $6\times 6$ mixing matrix $U_u^R$ takes an abnormal structure 
as if the third and fourth rows are exchanged. 
This is due to the top-quark-mass enhancement and the fourth 
up-quark-mass suppression in the model.

We have found that the fourth up-quark $t'$ is considerably light 
compared with the other heavy fermions ($m_{t'}\sim$ a few TeV  
in contrast to the other $m_F\sim $ a few hundred TeV) and it can couple 
to the right-handed weak boson $W_R$ with a sizable magnitude 
$|V_{t' d}^R|$, i.e., $|V_{t' d}^R|\simeq |V_{td}^L|$.
Therefore,  we can expect the observation of the single 
$t'$-production, $u+d\rightarrow d+t'$, at LHC.

We have investigated possible FCNC effects within the framework 
of SU(2)$_L\times$ SU(2)$_R\times$U(1)$_Y$ gauge model. 
We have estimated the cross section of the single top-quark 
production $e^+ e^- \rightarrow t \overline{c}$ through FCNC 
and obtained 
$\sigma(e^+ e^-\rightarrow t\overline{c}+ c\overline{t})
\simeq 1.1\times 10^{-4}$ pb at $\sqrt{s}=0.5$ TeV, 
so that the single top-quark production 
can be barely  detected at future $e^+ e^-$ colliders 
with  high luminosity such as JLC.

Thus, the exciting aspect of the present model is that the right-handed
fermion mixing matrix $U_u^R$ in the up-quark sector has a peculiar 
structure.  We hope that non-standard effects from such an abnormal 
structure of $U_u^R$ will be observed at future colliders such as 
JLC.

\vglue.2in

\centerline{\large\bf Acknowledgments}

A series of works based on the democratic seesaw mass matrix model
was first started in collaboration with H.~Fusaoka.
The author would like to thank H.~Fusaoka for his enjoyable 
collaboration. 
The author also thank R.~Hamatsu and A.~Miyamoto for information 
on the collider parameters of HERA and JLC, respectively.
This work was supported by the Grant-in-Aid for Scientific Research, the 
Ministry of Education, Science and Culture, Japan (No.08640386). 




\vglue.2in
\newcounter{0000}
\centerline{\large\bf References}
\begin{list}
{[~\arabic{0000}~]}{\usecounter{0000}
\labelwidth=0.8cm\labelsep=.1cm\setlength{\leftmargin=0.7cm}
{\rightmargin=.2cm}}
\item Y.~Koide and H.~Fusaoka, Z.~Phys. {\bf C71} (1966) 459.
\item Z.~G.~Berezhiani, Phys.~Lett.~{\bf 129B} (1983) 99;
Phys.~Lett.~{\bf 150B} (1985) 177;
D.~Chang and R.~N.~Mohapatra, Phys.~Rev.~Lett.~{\bf 58},1600 (1987); 
A.~Davidson and K.~C.~Wali, Phys.~Rev.~Lett.~{\bf 59}, 393 (1987);
S.~Rajpoot, Mod.~Phys.~Lett. {\bf A2}, 307 (1987); 
Phys.~Lett.~{\bf 191B}, 122 (1987); Phys.~Rev.~{\bf D36}, 1479 (1987);
K.~B.~Babu and R.~N.~Mohapatra, Phys.~Rev.~Lett.~{\bf 62}, 1079 (1989); 
Phys.~Rev. {\bf D41}, 1286 (1990); 
S.~Ranfone, Phys.~Rev.~{\bf D42}, 3819 (1990); 
A.~Davidson, S.~Ranfone and K.~C.~Wali, 
Phys.~Rev.~{\bf D41}, 208 (1990); 
I.~Sogami and T.~Shinohara, Prog.~Theor.~Phys.~{\bf 66}, 1031 (1991);
Phys.~Rev. {\bf D47} (1993) 2905; 
Z.~G.~Berezhiani and R.~Rattazzi, Phys.~Lett.~{\bf B279}, 124 (1992);
P.~Cho, Phys.~Rev. {\bf D48}, 5331 (1994); 
A.~Davidson, L.~Michel, M.~L,~Sage and  K.~C.~Wali, 
Phys.~Rev.~{\bf D49}, 1378 (1994); 
W.~A.~Ponce, A.~Zepeda and R.~G.~Lozano, 
Phys.~Rev.~{\bf D49}, 4954 (1994).
\item H.~Terazawa, University of Tokyo, Report No.~INS-Rep.-298 (1977) 
(unpublished); Genshikaku Kenkyu (INS, Univ.~of Tokyo) {\bf 26}, 33 
(1982);
Y.~Koide, Phys.~Rev. {\bf D49}, 2638 (1994).
\item M.~Gell-Mann, P.~Rammond and R.~Slansky, in {\it Supergravity}, 
edited by P.~van Nieuwenhuizen and D.~Z.~Freedman (North-Holland, 
1979); 
T.~Yanagida, in {\it Proc.~Workshop of the Unified Theory and 
Baryon Number in the Universe}, edited by A.~Sawada and A.~Sugamoto 
(KEK, 1979); 
R.~Mohapatra and G.~Senjanovic, Phys.~Rev.~Lett.~{\bf 44}, 912 (1980).
%
%
\item H.~Harari, H.~Haut and J.~Weyers, 
Phys.~Lett.~{\bf 78B}, 459 (1978);
T.~Goldman, in {\it Gauge Theories, Massive Neutrinos and 
Proton Decays}, edited by A.~Perlumutter (Plenum Press, New York, 
1981), p.111;
T.~Goldman and G.~J.~Stephenson,~Jr., Phys.~Rev.~{\bf D24}, 236 (1981); 
Y.~Koide, Phys.~Rev.~Lett. {\bf 47}, 1241 (1981); 
Phys.~Rev.~{\bf D28}, 252 (1983); {\bf 39}, 1391 (1989);
C.~Jarlskog, in {\it Proceedings of the International Symposium on 
Production and Decays of Heavy Hadrons}, Heidelberg, Germany, 1986
edited by K.~R.~Schubert and R. Waldi (DESY, Hamburg), 1986, p.331;
P.~Kaus, S.~Meshkov, Mod.~Phys.~Lett.~{\bf A3}, 1251 (1988); 
Phys.~Rev.~{\bf D42}, 1863 (1990);
L.~Lavoura, Phys.~Lett.~{\bf B228}, 245 (1989); 
M.~Tanimoto, Phys.~Rev.~{\bf D41}, 1586 (1990);
H.~Fritzsch and J.~Plankl, Phys.~Lett.~{\bf B237}, 451 (1990); 
Y.~Nambu, in {\it Proceedings of the International Workshop on 
Electroweak Symmetry Breaking}, Hiroshima, Japan, (World 
Scientific, Singapore, 1992), p.1.
\item N.~Cabibbo, Phys.~Rev.~Lett.~{\bf 10}, 531 (1996); 
M.~Kobayashi and T.~Maskawa, Prog.~Theor.~Phys.~{\bf 49}, 652 (1973).
\item Y.~Koide,  Mod.~Phys.~Lett.{\bf A36}, 2849 (1996).
\item Y.~Fukuda {\it et al.}, Phys.~Lett. {\bf B335}, 237 (1994).
\item J.~C.Pati and A.~Salam, Phys.~Rev. {\bf D10}, 275 (1974); 
R.~N.~Mohapatra and J.~C.~Pati, Phys.~Rev. {\bf D11},
 366 and 2588 (1975);
G.~Senjanovic and R.~N.~Mohapatra, Phys.~Rev. {\bf D12}, 1502 (1975).
%
\item Y.~Kimura, presented at The 5th ICFA Seminar at KEK {\it ``Future 
Perspectives in High Energy Physics"}, Oct. 15-18, 1996, to be published in 
the Proceedings.
\end{list}

\end{document}